\documentclass[12pt, twocolumn]{emulateapj}

\usepackage{longtable}

\usepackage[usenames]{color}
\usepackage{ulem}
\usepackage{tablefootnote}

\shorttitle{X-ray from SBs N\,70 and N\,185}
\shortauthors{Reyes-Iturbide et al.}

\begin{document}

\title{Diffuse X-ray emission from the superbubbles N\,70 and N\,185 in
  the Large Magellanic Cloud
    }

\author{Reyes-Iturbide J.\altaffilmark{1}, Rosado,
  M.\altaffilmark{2},  Rodr\'\i
  guez-Gonz\'alez, A.\altaffilmark{1},
 Vel\'azquez  P. F. \altaffilmark{1}, S\'anchez-Cruces
 M.\altaffilmark{2}, and Ambrocio-Cruz, P\altaffilmark{3}.} 

\altaffiltext{1}{Instituto de Ciencias
Nucleares, Universidad Nacional Aut\'onoma de M\'exico, Apdo. Postal
70-543, D.F.  M\'exico}

\altaffiltext{2}{Instituto de Astronom\'\i a,
  Universidad Nacional Aut\'onoma de M\'exico, Apdo. Postal 70-264,
  C.P. 04510, M\'exico, D.F., M\'exico}

\altaffiltext{3}{Instituto de Ciencias B\'asicas e Ingenier\'\i a, Universidad Aut\'onoma del Estado de Hidalgo, Ciudad Universitaria, Km 4.5 Carretera Pachuca-Tulancingo; Col. Carboneras, C.P. 42184, Mineral de la Reforma, Hgo., M\'exico.}

\begin{abstract}

We present a study of the diffuse X-ray emission from superbubbles
N\,70 (DEM L301) and N\,185 (DEM L25) located in the Large Magellanic
Cloud, based on data from the XMM-Newton Satellite. We obtained
spectra and images of these objects in the soft X-ray energy band. These 
 X-ray spectra were fitted by a thermal plasma model, with temperatures of $2.6
\times 10^{6}$ K and $2.3 \times 10^{6}$ K,  for N\,70 and N\,185,
respectively.
For N70, images show that X-ray emission comes from the
inner regions of the superbubble, when we compare the distribution of
the X-ray and the optical emission; while for N\, 185, the X-ray
emission is partially confined by the optical shell. We suggest that
the observed X-ray emission is caused by shock-heated gas, inside of
the optical shells. We also obtained  X-ray luminosities which exceed
the values predicted by the standard analytical model. This fact shows
that, in addition to the winds of the interior stars , it is necessary
to consider another ingredient in the description, such as a supernova
explosion, as has been proposed in previous numerical models.

\end{abstract}


\section{Introduction}

Massive stars transfer energy into the interstellar medium (ISM) 
in two ways during their lifetime: by radiative luminosity ($L_{\star}$)
and by mechanical luminosity due to winds ($L_{\rm w}$)--- where $L_{\rm w}= \frac {1}
{2}\dot M_{\rm w} \rm v_{\rm w}^2$, $\dot M_{\rm w}$ is the mass loss rate and $ \rm v_{\rm w}$ is
the wind terminal velocity---. In the ``single scattering upper
limit'' approximation, which assumes that the momentum of the 
wind is equal that of the radiation, the mechanical luminosity of the
stellar wind is less than one per cent of the stellar radiative
luminosity, $L_{\rm w}/L_{\star}=\frac {1}{2} \rm v_{\rm w}/c \lesssim 0.002$,
for a typical terminal velocity of winds from early-type stars 
(corresponding to $\rm v_{\rm w}=1000$ km s$^{-1}$ and 
$\dot M_{\rm w}=10^{-6}M_{\odot}$~yr$^{-1}$), and $c$ is the speed of light. However, 
mechanical luminosity transfers energy to the interstellar medium
(ISM) more efficiently. 
A classical example is found in OB 
associations (OBAs), which are composed of early-type massive stars that contain 
strong stellar winds, which in turn produce shock waves that sweep up
the surrounding ISM, creating a superbubble (SB) around the OBA.
The standard model \citep{w77} describes these SBs as structures that
consist of a shell of swept-up ISM, cool and bright in optical
emission lines, that contains shock-heated gas emitting X-rays in its
interior. 

The standard model has been tested with observations of the kinematics and
the X-ray emission of bubbles and superbubbles. 
Of particular relevance are the studies of SBs in the Large Magellanic Cloud (LMC) that have
large-diameter shells, which in several cases are larger than
predicted by the standard model.  \cite{chum90} and \cite{wh91} found
that most SBs had X-ray luminosities an order of
magnitude higher than predicted by this model. Nevertheless,
there are superbubbles with luminosities consistent with the predictions
of the standard model. \cite{oey96b} --based on
observations by \cite{ros81}, \cite{ros82} and \cite{ros86}-- proposed
two SB categories in terms of dynamical data: high-velocity and low-velocity
superbubbles, latter type being more consistent with the standard
model. \citet{oey96b} concluded that X-ray emitting SBs with
high expansion velocities and intermediate [\ion{S}{2}]/[H $\alpha$]
line-ratios ($\geq 0.5$) cannot be explained by the mechanical
energy of stellar winds alone, and must be associated with supernova
explosions that occurred in their interiors. The energy released by the
supernova explosion would be an additional source of heating for the
gas inside the superbubble, which would explain the observed X-ray
excess. At the same time, the explosion would produce an acceleration of
the shell that could explain the high expansion velocities
observed.  In a previous  work \citep{ary11}  it has been confirmed
the high expansion velocities in the models in which a supernova event
has taken place.
For instance, the Gum Nebula is a case in which a supernova explosion
can explain the X-ray emission  \citep{lea92}. However, as discussed
in \cite{ros86} the detection of non-thermal radio emission is
essential to have a definitive confirmation of the presence of a
supernova remnant (SNR) in the SB. 

In this work we study two SBs belonging to
the high-velocity type: the N\,70 and N\,185 (both in the LMC). This
characteristic makes these two superbubbles provide excellent laboratories
in which to study the X-ray emission.

The superbubble N\,70 (DEM L301) is an almost circular shell,
about $100$ pc in diameter. It contains  the
OBA LH~114 \citep{lh70}, with more than $1000$ stellar
sources (The SIMBAD Astronomical Database); of these, seven are O-type
stars \citep{oey96a}.  The mean age of this star association is
about $5$ Myr, a sufficient time for the first supernova
explosion to occur. While radio observations are inconclusive about
the non-thermal character of its radio continuum emission
\citep{mil80}, a subsequent work by \citep{fil98} reported a spectral
index  $\alpha=-0.02$ which is larger than the typical $\alpha=-0.43$
value for SNRs.
Observations in the optical and X-ray bands have 
found evidence of a supernova explosion inside the superbubble
\citep{ros81}. \cite{ros81} and \cite{george83} obtained intermediate
[\ion{S}{2}]/[H$\alpha$] line-ratios in N\,70, larger than those
typical of photoionized HII regions, but lower than those
typical of SNRs in the LMC. The range of values ​found in the outer filamentary
shell  is ($[0.8-2.0]$), are close to the range of excitation in some
of the supernova remnants observed in the LMC \citep{sk99}.
These line ratios can be explained by models with shock velocities of
about $40-70$ km~s$^{-1}$, in agreement with the measured expansion
velocity of this superbubble. The high expansion velocity is in contradiction
with the standard model (Oey et al. 1996b).
Using archival data from the Einstein observatory, \cite{chum90} 
reported an X-ray luminosity of N\,70 an order of
magnitude higher than predicted by the classical wind-blown bubbles
model.

The superbubble N\,185 (DEM L25) is very similar to N\,70. N\,185 has a
spherical shape, with a diameter between $92$ and $112$ pc. \cite{ros82}
measured an expansion velocity of $70$ km~s$^{-1}$ for this
superbubble.  N\,185 contains an OB association in its interior,
consisting of more than 800 stellar sources (The
SIMBAD Astronomical Database);  where only one is an O-type star
\citep{oey96a}. The likelihood that it had some additional massive stars in
the past, along the high expansion velocity, and the intermediate
[\ion{S}{2}]/[H $\alpha$] line-ratio \citep{ros82, george83}  suggest
that this superbubble could have originated from a supernova
explosion, or explosions. 
Radio observations have reported a spectral index $\alpha=-0.67$,
which is typical of SNR candidates \citep{fil98}.  N\,185 was detected
by ROSAT, with an X-ray luminosity of  $L_{\rm X}\sim 1.8 \times
10^{35}$~erg~s$^{-1}$ in \cite{oey96a}; this is higher than what is
expected from standard model predictions.

To unravel the origin of these high-velocity superbubbles, we will
focus our attention to their X-ray emission, now observed with 
more sensitive instruments such as XMM-Newton observatory. We selected
these two SBs because of their high expansion velocities, their
high [\ion{S}{2}]/[H $\alpha$] line-ratios, and their X-ray excess.  A
detailed study of the X-ray emission could help us to determine if
stellar winds alone are the origin of the large diameter shells, or if
supernovae are required to produce the large shells. If supernova
explosions have occurred in their interiors, then the X-ray emission
might allow us to determine the relative contributions to the bubble
expansion from the stellar winds and from the supernova explosions.

Our study of diffuse X-ray emission from SBs N\,70 and N\,185
is made with archival data from XMM-Newton observations. This
paper is organized in the following way: In Section 2 we present
 a brief overview to the theory of superbubbles. Section 3 we
 present the data and describe its reduction. Sections 4 and 5 are
 dedicated to  the observational results  for the SBs N\,70 and
 N\,185, respectively. Finally, in Section 6 we give a discussion and
 our conclusions.

\section{Superbubble dynamics and soft X-ray emission}

The standard model used to describe the structure of SBs is based on
the original model of \citep{w77}, which describes the structure
produced by the interaction of a wind of a single star with its
environment, and it was later extended to include the winds of several
stars in a OBA   \citep{chu95}.
The mechanical energy that the stars of the OBA deposit to the ISM in the
form of stellar winds, is given by 

\begin{equation}
\label{eq1}
L_{\rm w}=\sum_{i=1}^{N}\frac{1}{2}\dot M_{{\rm w},i} \rm v_{{\rm w},i}^2
\;,
\end{equation}
where $\dot  M_{{\rm w},i}$ and $\rm v_{{\rm w},i}$ are the mass-loss rate and the wind terminal velocity of the $i$th star, respectively, and $N$ is the total number of stars.
The interaction of this winds with ISM creates a superbubble structure
with the following four regions (see Figure \ref{fig1}): 

%%%%%%%%%%%%%%%%%%%%%%%%%%%%%%%%%%%%%%%%%%%%%%%%%%%%%%%%%%%%%%%%%%%%%%
\begin{figure}
\begin{center}
\includegraphics[angle=0,scale=0.4]{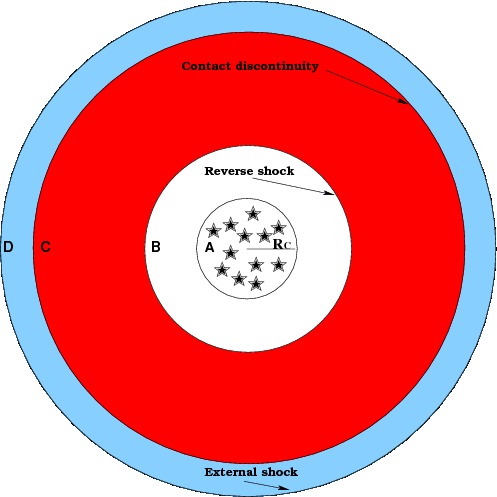}
\caption{A generalization of the standard model \citep{w77} conceived for a single star. Schematic structure of a superbubble produced by an OBA :
  (A) star cluster region, (B) free-wind region, (C) shocked-wind
  region, and (D) shocked-ISM region. 
\label{fig1}}
\end{center}
\end{figure}
%%%%%%%%%%%%%%%%%%%%%%%%%%%%%%%%%%%%%%%%%%%%%%%%%%%%%%%%%%%%%%%%%%%%%%

\begin{itemize}
\item (A) An inner zone where the stars are located and inject their
  winds. This region is delimited by the OBA radius $R_{\rm
    C}$. Outside this region a common OBA wind is established.

\item (B) A free-wind zone is the region between the OBA radius and
  the reverse shock.This region is filled by the unperturbed stellar
  OBA wind.

\item (C) A shocked OBA wind zone located between a reverse shock
  (or inner shock) and the contact discontinuity. 

\item (D) An external zone, between the contact discontinuity and the
  external shock. This region contains shocked ISM material that has
  been swept by the external shock. 

\end{itemize}

In zone A, stellar winds of the massive stars collide with each other,
thermalizing all the gas injected inside the cluster volume forming a
common OBA wind --for this to be possible $R_{\rm C} \ll R$, where $R$
is the radius of superbubble--.
The OBA wind expands freely inside zone  B.
Zone C is formed by gas of the stellar cluster wind that has been
shocked by the inner or reverse shock. This outflow is thermalized and
emits in soft X-ray energy range. Finally, zone D is formed by 
shocked-ISM gas emitting in optical; it is the densest zone according
to the standard model. This description corresponds to the
intermediate stage of evolution of the superbubble, there are
important radiative losses in the region (D) but in region (C) there
are small; and in the contact discontinuity that separates them
thermal conduction transports heat  due to the large temperature
gradient between the regions (C) and (D) which have
temperatures of $10^{6}$ and $10^{4}$, respectively. 

The equations that described the dynamics of the shell or region (D)
according to  the \citet{w77} model are:

\begin{equation}
\label{eq2}
R=(42 \, {\rm pc})L_{\rm w37}^{1/5}n_{\rm 0}^{-1/5}t_{\rm 6}^{3/5}
\;,
\end{equation}

\begin{equation}
\label{eq3}
V=\frac{dR}{dt}=(0.59 \, {\rm km \, s}^{-1}) R_{\rm pc}/t_{\rm 6}
\;.
\end{equation}
$R$ and $V$ are the radius and expansion velocity of the SB; where
$R_{\rm pc}$ is the radius in units of pc, $L_{\rm w37}$ is the
mechanical luminosity of the OBA in units of $10^{37}$ erg s$^{-1}$,
$n_{\rm 0}$ and $t_{\rm 6}$ are the number density of the ambient
medium in units of cm$^{-3}$ and the age of the bubble in
$10^{6}$ years, respectively. 

 The X-ray luminosity of a spherically symmeric gas distribution is
 given by the integral X-ray emissivity in the layer of shocked
 stellar wind --region (C)--:  

\begin{equation}
\label{eq4}
L_{\rm X}= \int n^{2}(R)\Lambda_{\rm X}(Z,T)d^{3}R
\;,
\end{equation}
where $R$ is the radial coordinate, $n(R)$ is the numerical density,
and $\Lambda_{\rm X}(Z, T)$ is the X-ray emissivity as a function of
its temperature and metallicity. The X-ray luminosity that arises from
the shocked gas (zone C) in the superbubble evolving inside a homogeneous
ISM can be estimated by integrating Eq (4) from center to radius
$R_{\rm max}$ where the temperature drops to the minimum (i.e.,
$T(R_{\rm max})=T_{\rm min}$). As presented by \cite{w77} and
\cite{chu95}, the integral can be written as:

\begin{equation}
\label{eq5}
L_{\rm X}=1.1\times 10^{35}I(\tau)\xi L_{\rm w37}^{33/35}n_{\rm
  0}^{17/35}t_{\rm 6}^{19/35} ({\rm erg~s}^{-1})\;,
\end{equation}
where $\xi$ is the gas metallicity, and
\begin{equation}
\label{eq6}
I(\tau)=\frac{125}{33}-5\tau^{1/2}+\frac{5}{3}\tau^{3}-\frac{5}{11}\tau^{11/3} 
\;, 
\end{equation}
with
\begin{equation}
\label{eq7}
\tau=0.16L_{w37}^{-8/35}n_0^{-2/35}t_6^{6/35}
\;.
\end{equation}

The expression for the X-ray luminosity --Equation(\ref{eq5})-- is
practical because $L_{x}$ can be derived from physical parameters
obteined from observations such as ISM density, expansion velocity and the
size of the SB, and the mechanical energy of the OBA wind. 

\section{XMM-Newton data and data reduction}

Superbubbles N\,70 and N\,185 were observed with The European Photon
Imaging Ca\, mera (EPIC), on board of the XMM-Newton observatory. EPIC
consists of two Metal Oxide Semi-conductor (MOS) CCD arrays (MOS1 \&
MOS2, \citealt{turn21}) and a pn-CCD \citep{stru21}.  A summary of the
observations is given in Table~\ref{tab1}. The EPIC MOS cameras were
operated in the Prime Full-Window Mode and the EPIC pn camera was
operated in the Extended Prime Full-Window Mode. Two of the cameras
contain seven MOS CCDs, while the third uses twelve PN CCDs, defining
a circular field of vision (FOV) of $\sim 30 \arcmin$ in diameter,
allowing the inclusion  of the entire superbubbles (sizes between
$6\arcmin-10\arcmin$) in one pointing.
In both observations, a medium filter was used to block
ultraviolet photons (\cite{vill98} and \cite{ste96}). The XMM-Newton
pipeline products were processed using the XMM-Newton Science Analysis
Software (SAS version 11.0.0).

Since the emission from these superbubbles is extended, the data
were processed with the XMM-Newton Extended Source Analysis software
package (XMM-ESAS) \citep{snow08} for the analysis of EPIC MOS and pn
observations. XMM-ESAS documentation can be found at the SAS Package
documentation web pages: {\it http://xmm.esac.esa.int/sas/ \, \,
  current/howtousesas.shtml}. This package automatically filters times
of high background contamination. The resulting effective exposure
times are given in Table~\ref{tab1}.

After filtering the event files, we created mosaic images of the EPIC
cameras. We used the ``cheese'' task to identify points sources and
remove them. The package includes tasks to create and model the
non-cosmic background, correct the exposures and subtract the
backgrounds. 
We generate combined images (with a pixel size of 2.5$\arcsec$) of the
EPIC cameras in the $([0.4-1.5])$~ keV bands; there is no significant
emission of the diffuse X-ray emission above 2.0 keV for both
superbubbles. Finally, they were smoothed as described in Section 4.1
and 5.1.

A good background knowledge is crucial for the spectral analysis.  For
a point source, a local background extracted from a neighboring region
can be used. In the case of a diffuse source, as in the case of SBs
N\, 70 and N\,185 corresponding to an area of diffuse X-ray emission
that covers one third of the XMM-Newton FOV --filling entirely CCD\#1
or the central CCD of EPIC MOS--, it is inappropriate to estimate the local
background from the same data; many effects can be produced due to
differences in chip position. The high-energy particles that interact
with material surrounding the detector produce fluorescence, which
varies with position on the detector, especially for the PN
detector. In addition, the spectral response depends on the position
on the detector.  To estimate the background of diffuse sources, the
XMM-Newton EPIC Background working group has created the so called
blank sky data for each EPIC \citep{cart07}. The blank sky data have
been merged with data from different pointings after the point sources
were eliminated. This data set consists in the detector
background and an average cosmic X-ray background. In this paper we
used the blank sky data obtained by XMM-Newton EPIC Background
Working Group (Request id: 0573-0575) in the direction of the SBs
N\,70 and N\,185.

Before extracting the spectra, we have first corrected both the
observed data and the blank sky data for vignetting using the XMMSAS
command evigweight.  The spectra of the diffuse X-ray emission from
N\,70 and N\,185 were extracted from the event files of the EPIC
cameras. These spectra were extracted from a circular region
encompassing the entire superbubble, while the background spectrum is
extracted from the blank sky data at the same location on the detector
as the source spectrum. For this we use the script skycast, which
converts the detector coordinates of the XMM EPIC background into sky
coordinates, using the pointing direction of the observation. Finally,
the spectra are grouped with a minimum of 25 counts/bin.

\section{Results for N\,70}

\subsection{Images: X-ray brightness distribution} 

%%%%%%%%%%%%%%%%%%%%%%%%%%%%%%%%%%%%%%%%%%%%%%%%%%%%%%%%%%%%%%%%%%%%%%
\begin{figure*}
\begin{center}
\includegraphics[angle=0,scale=.435]{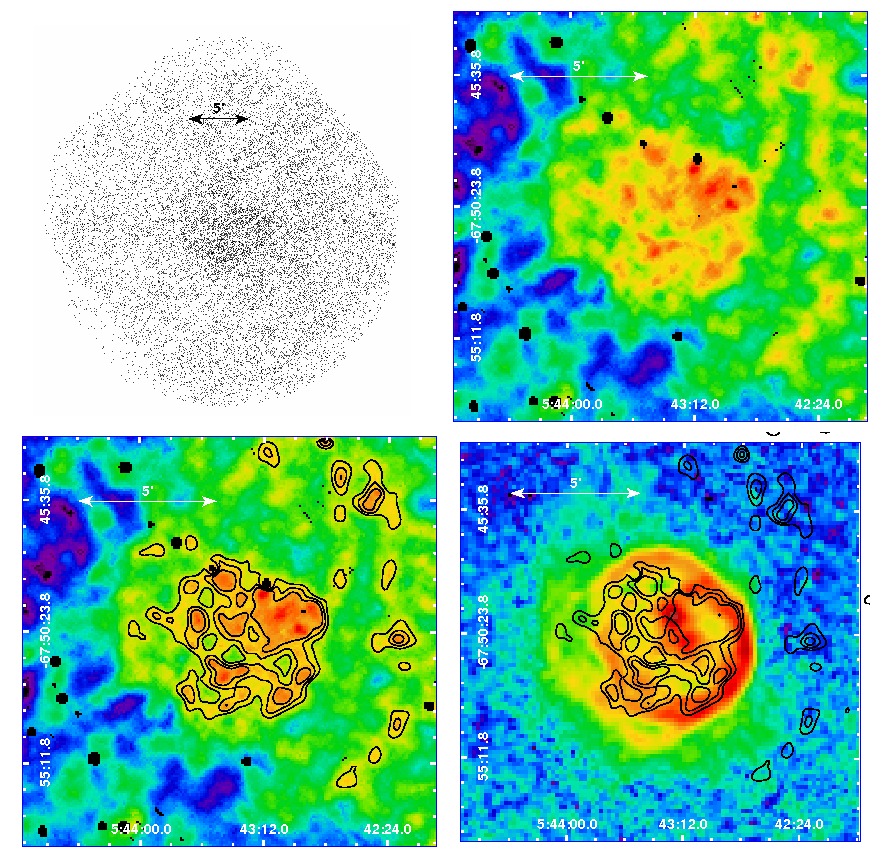}
\caption{The mosaic EPIC images of N\, 70, in the ($[0.4-1.5]$) keV
  energy band. Top row: left, the raw data, combined EPIC MOS1/2 and
  pn image; right, the combined EPIC smoothed image. Bottom row: left,
  X-ray contours at $3\sigma$, $4\sigma$ and $5\sigma$ above the
  cosmic background level are superposed onto the smoothed image; right, a
  comparison between the H$\alpha$ emission of N\,70, which was taken
  from the Magellanic Cloud Emission Line Survey (MCELS; Smith et
  al. 1998), and the X-ray emission (in black contours). Note the
  X-ray emission observed to the West, it is external to the optical
  superbubble, which is related to some field sources detected in
  this observation.
\label{fig2}}
\end{center}
\end{figure*}
%%%%%%%%%%%%%%%%%%%%%%%%%%%%%%%%%%%%%%%%%%%%%%%%%%%%%%%%%%%%%%

\begin{figure}
\begin{center}
\includegraphics[angle=0,scale=0.7]{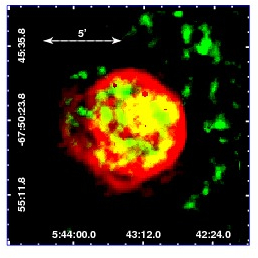}
\caption{ RGB image of N\,70.  MCELS image of H$\alpha$ emission is
  shown in red, while the diffuse soft X-ray emission is displayed in
  green.
\label{fig3}}
\end{center}
\end{figure}

Figure \ref{fig2} shows a mosaic of images in the $([0.4-1.5])$~ keV
energy band in the direction of SB N\,70.  The top-left panel
displays a combined EPIC MOS1/2 and pn FOV image, which have been
obtained by using the task {\it comb}. The top-right panel shows the
combined EPIC smoothed (with a kernel of $80$ counts) image, with a size
of $15\arcmin \times 15\arcmin$, where the non-cosmic background was
subtracted and the exposure has been corrected. Besides, several point
sources have been identified and removed from this observation, which
are indicated by small circular voids in this image. A $15 \arcmin$
square smoothed image is displayed in the bottom-left panel of
Fig. \ref{fig2}, which is overlayed by the X-ray contours at
$3\sigma$, $4\sigma$ and $5\sigma$ above the cosmic background
level. In order to analyze the spatial distribution of the soft X-ray
emission, in the bottom-right panel we show a comparison between the
H$\alpha$ emission of N\,70, which was taken from the Magellanic Cloud
Emission Line Survey (MCELS; Smith et al. 1998), and the X-ray
emission (in black contours). We note that the soft X-ray emission is
well-confined within the H$\alpha$ shell. This spatial distribution
suggests that the X-ray emission is produced by hot gas inside
the optical superbubble. This distribution is recreated with an RGB
image--panel (Figure \ref{fig3}) where the H$\alpha$ emission is
displayed in red and the diffuse soft X-ray is shown in green. Note
the emission to the East, external to the optical superbubble, it is
related to some sources found in the field of vision.

%%%%%%%%%%%%%%%%%%%%%%%%%%%%%%%%%%%%%%%%%%%%%%%%%%%%%%%%%%%%%%%%%%%%%%%%%%
%%%%%%%%%%%%%%%%%%%%%%%%%%%%%%%%%%%%%%%%%%%%%%%%%%%%%%%%%%%%%%%%%%%

\subsection{Spectra and X-ray luminosity from diffuse emission}

\begin{figure}
\includegraphics[angle=-90,scale=0.35]{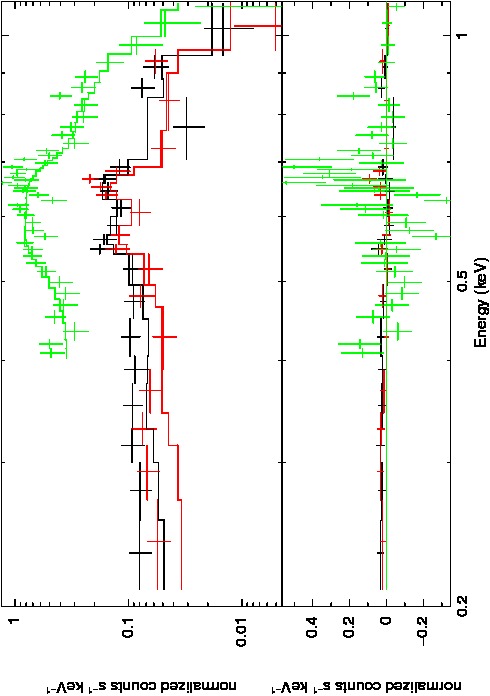}
\caption{XMM-EPICs background subtracted source spectra of the diffuse X-ray emission from N70. The solid lines show the best-fit model. In black, red and green EPIC MOS1, MOS2 and PN, respectively.
\label{fig4}}
\end{figure}

For spectral analysis of the EPIC MOS data, we used XSPEC version
12.5.0 (distributed with the HEASsoft 6.6.1 software). The spectra of
the EPIC MOS1/2 and EPIC pn were extracted from a circular region
encompassing entirely the N\,70 superbubble. The point sources inside
this region were excluded. The background contribution was estimated
from blank sky data for each EPIC-MOS1/2 and EPIC-PN; these contain
both contributions (i.e., instrumental and cosmic). The background
subtracted source spectra are shown in Figure \ref{fig4}. We present
the EPIC MOS1/2 and pn spectra in the interval of  energy
($[0.2-1.1]$) keV and ($[0.4-1.1]$), respectively.

%%%%%%%%%%%%%%%%%%%%%%%%%%%%%%%%%%%%%%%%%%%%%%%%%%%%%%%%%%%%%%%%%%%%%%%%%
%%%%%%%%%%%%%%%%%%%%%%%%%%%%%%%%%%%%%%%%%%%%%%%%%%%%%%%%%%%%%%%%%%%%%%%%%

In order to obtain the physical conditions of the X-ray-emitting gas,
the EPIC spectra were fitted simultaneously with a thermal model
(APEC) \citep{smith01} convolved with the interstellar absorption. The
best fit gives a reduced $\chi^{2}=1.3$
($\chi^{2}/d.o.f=320.6/242$). We used a single absorbing column
density via the photoelectric absorption model \citep[PHABS][]{bal92},
using reasonable values for the absorption column 
density, $N_{H}=5.0 \times 10^{20}$ cm$^{-2}$, which is in
agreement with the measures of column densities in the LMC direction,
average $N_{H}=6.4 \times 10^{20}$ cm$^{-2}$ \citep{dic90}. The
chemical abundance was set to $0.3$ times the solar abundance, i.e.,
the average value of the ISM in the LMC \citep{rus92,hug98}. This
best-fit has plasma temperatures of $kT=0.22$~keV. The observed total
flux of the diffuse X-ray emission, corrected for absorption, is $
(0.7 \pm 0.1) \times 10^{-12}$~erg~cm$^{-2}$~s$^{-1}$ in the
$[0.2-1.1]$ keV range, which corresponds to X-ray luminosity (at the
LMC distance of $54$~kpc, \citet{fea99}) of $2.4 \pm 0.4 \times
10^{35}$~erg~s$^{-1}$. The best-fit model is overplotted on the EPIC
spectra with a solid curves in Figure \ref{fig4}, and the best-fit
parameters are listed un Table~\ref{tab2}.

\section{Results for N\,185}

\subsection{X-ray brightness distribution}

%%%%%%%%%%%%%%%%%%%%%%%%%%%%%%%%%%%%%%%%%%%%%%%%%%%%%%%%%%%%%%%%%%%%%%%%%%%%%%%%
\begin{figure*}
\begin{center}
\includegraphics[angle=0,scale=.435]{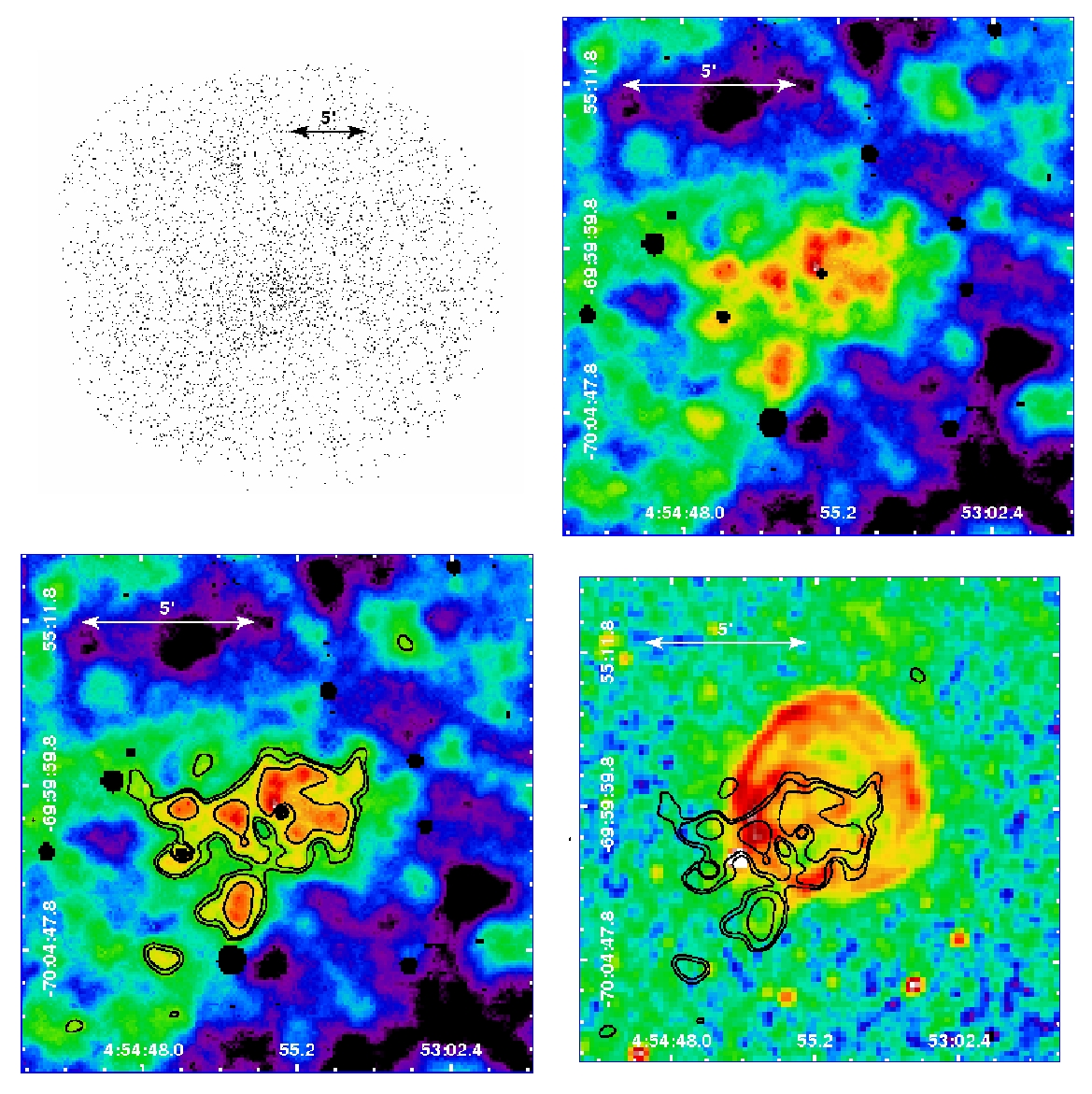}
\caption{The mosaic EPIC images from N\, 185, in the $([0.4-1.5])$ keV
  energy band. Top row: left, the raw data, combined EPIC MOS1/2 and
  pn image; right, the combined EPIC smoothed image. Bottom row: left,
  X-ray contours at $5\sigma$, $6\sigma$ and $9\sigma$ above the
  cosmic background level are superposed onto smoothed image; a
  comparison between the H$\alpha$ emission of N\,185, which was taken
  from the MCELS (Smith et
  al. 1998), and the X-ray emission (in black contours).
\label{fig5}}
\end{center}
\end{figure*}
%%%%%%%%%%%%%%%%%%%%%%%%%%%%%%%%%%%%%%%%%%%%%%%%%%%%%%%%%%%%%%%%%%%%%%%%%%%%

Figure \ref{fig5} shows an image mosaic of N\,185, in the
$([0.4-1.5])$~ keV energy band, which is similar to that shown in
Figure \ref{fig2} for the case of N\,70. The combined EPIC MOS1/2 and pn FOV
image is displayed in the top-left panel while in the top-right panel
is shown the combined EPIC smoothed image (in this case with a kernel
of 50 counts). The bottom-left panel show the smoothed X-ray image,
which is overlayed by X-ray contours at $5\sigma$, $6\sigma$ and
$9\sigma$ above the cosmic background level. Finally, the comparison between
H$\alpha$ (color-scale) and X-ray (black contours) emission is shown in
the bottom-right panel. As in Figure \ref{fig2}, the H$\alpha$ emission
was taken from the Magellanic Cloud Emission Line Survey (MCELS; Smith
et al. 1998).

We note that, in the case of N\,185, the soft X-ray emission is only
partially confined by the H$\alpha$ shell, X-ray emission extends well
beyond the optical shell to  the southeast direction, where the
optical shell appears bright and well defined, so it is probable that
the gas is escaping towards the 
back of the SB. 

\begin{figure}
\includegraphics[angle=0,scale=0.8]{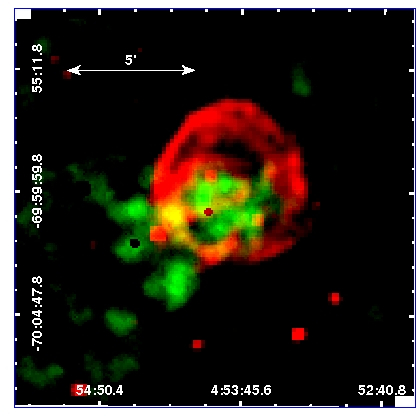}
\caption{ RGB image of N\,185. As in Figure \ref{fig5} the MCELS
  H$\alpha$ and the diffuse soft X-ray
  emissions are  displayed in red and green, respectively.
\label{fig6}}
\end{figure}

This comparison  is recreated with an RGB image (Figure
\ref{fig6}) where H$\alpha$ and soft X-ray emission are displayed
in red and green, respectively. 

%%%%%%%%%%%%%%%%%%%%%%%%%%%%%%%%%%%%%%%%%%%%%%%%%%%%%%%%%%%%%%%%%%%%%%%%%%%%
%%%%%%%%%%%%%%%%%%%%%%%%%%%%%%%%%%%%%%%%%%%%%%%%%%%%%%%%%%%%%%%%%%%%%%%%%%

\subsection{X-ray spectra and luminosity from diffuse emission}

\begin{figure}
\begin{center}
\includegraphics[angle=-90,scale=0.35]{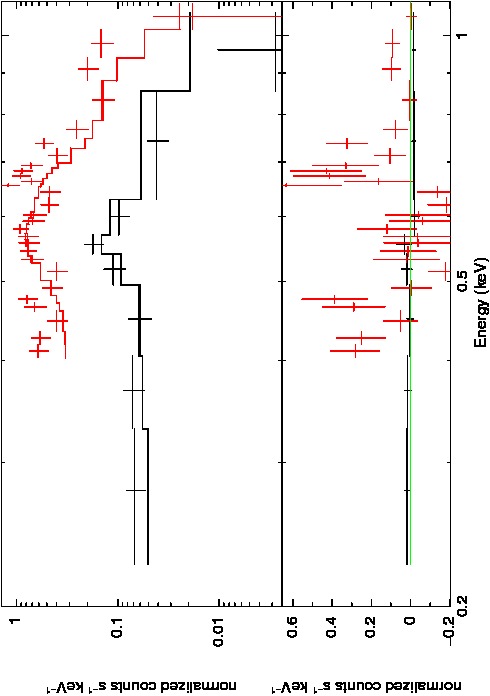}
\caption{XMM-EPICs background subtracted source spectra of the diffuse X-ray emission from N\,185. The solid lines show the best-fit model. In black  EPIC MOS1
 and PN in green.
\label{fig7}}
\end{center}
\end{figure}

The same method was employed to extract the spectra of the diffuse
emission and make the analysis of the superbubble N\,185. The
spectra of the EPIC MOS1/2 and EPIC PN were extracted from a circular
region encompassing the entirety of N\, 185 superbubble. The point
sources in this region were excluded from the analysis. The background
contribution was estimated from blank sky data for each EPIC-MOS1/2
and EPIC-PN data. The background subtracted source spectra are shown
in Figure \ref{fig7}. We present the EPIC MOS1 and pn spectra in the
interval of  energy ($[0.2-1.1]$) keV and ($[0.4-1.1]$),
respectively. Absorbed APEC model was used to fit this
spectrum. Figure~\ref{fig7} displays both the observed spectra and the
best-fit model spectrum.  To obtain this best-fit model, with a
reduced $\chi^{2}=1.6$, we have set the chemical abundance at the
LMC average values (i.e., $0.3$ times solar abundance
\citep{rus92,hug98}), and we used a column density value $N_H=5.0
\times 10^{20}$~cm$^{-2}$ in agreement with previous measurements of
column densities in the LMC direction \citep{dic90}. Plasma
temperatures of $kT=0.20$~keV, the observed total flux of the diffuse
X-ray emission, corrected for absorption, is $ (0.6 \pm 0.1) \times
10^{-12}$~erg~cm$^{-2}$~s$^{-1}$ in the $[0.2-1.1]$ keV range (see
Table \ref{tab3} ), which corresponds to X-ray luminosity (at the LMC
distance of $54$~kpc, \citet{fea99}) of $2.1 \pm 0.7 \times
10^{35}$~erg~s$^{-1}$.

%%%%%%%%%%%%%%%%%%%%%%%%%%%%%%%%%%%%%
%%%%%%%%%%%%%%%%%%%%%%%%%%%%%%%%%%%%%%%%%%%%%%%%%%%%%%%%%%%%%%%%%%%%%%%%%

\section{Discussion and conclusions}

We conducted a detailed study based on XMM-Newton data of the SBs
N\,70 and N\,185, in the LMC.  We have generated both X-ray images and
spectra of these objects. In both SBs, we find soft diffuse X-ray
emission. Soft X-ray emission comes from the inner region of the SBs
(i.e., interior to the optical shells).  Our results show thermal
spectra from SBs which are associated with the soft X-ray emission.

As discussed in Section 2, there is an analytical model that aims to explain
the soft X-ray emission from bubbles and superbubbles \citep{w77}, and has
become the standard model for SBs. According to this model, the supersonic
winds from massive (OB stars) interact with the interstellar medium
forming an expanding cold shell ($\sim 10^{4}$~K) with a hot interior
($\sim 10^{6}$~K) of shocked wind. We have calculated the X-ray
luminosity from superbubbles N\,70 and N\,185, using the Eq.
(\ref{eq5}). In this calculation we only take into account the stars
that dominate the mechanical luminosity of the stellar cluster
associated with these superbubbles
\citep[see][]{ros81,ros82,oey96a}.
Using the high terminal velocities and mass loss rates 
of stars with similar spectral types and luminosities, and considering
the average observational parameters for both superbubbles: an 
expansion velocity of $70$ km s$^{-1}$, SB radius of $50$ pc, and
ambient density of $0.1$ cm$^{-3}$  (from Eq. \ref{eq5}), the soft X-ray
luminosity is $6\times$10$^{34}$~erg~s$^{-1}$ for N\,70 and
$5\times$10$^{33}$~erg~s$^{-1}$ for N\,185. From our observations
we obtain an X-ray luminosity of $2.4\times$10$^{35}$~erg~s$^{-1}$
for N\,70, and of $2.1\times$10$^{35}$~erg~s$^{-1}$ for N\,185 (see
Table \ref{tab4}). 
Therefore, the X-ray luminosity predicted by the standard model is four times
lower than the observed values for N\,70, and $40$ times lower than
observed for N\,185. In this cases the standard model of wind-blown bubbles
cannot explain the observed soft X-ray emission of these SBs.

In a previous work, \citet{ary11} carried out numerical models of
the SB evolution applied to N\,70. As does the standard model, their
models predict that the primary X-ray emission will be in the soft
X-ray band and that it has a thermal origin ($\sim 10^6$~K).  In that work
several models were performed in order to reproduce the soft X-ray luminosity,
the high expansion velocity and the large radius of this
superbubble. The model including both a supernova explosion (absent in
the standard model of \citealt{w77}) and the
stellar winds fits quite well with the kinematics of the optical
shell,  and the excess of thermal X-ray emission of N\,70. An X-ray
luminosity of $ 2\times10^{35}$~erg~s$^{-1}$ was obtained for this model,
a value that agrees with the observed X-ray luminosity obtained for
N\,70 in this paper (see Table \ref{tab4}).

For N\,185, the same arguments can be employed to justify the
inclusion of a supernova, and indeed, non-thermal radio emission has
been detected as mentioned in the Introduction. A supernova
explosion is also consistent with the stellar population
models for N\,70 and N\,185 presented by \cite{oey96b}, in which 13
and 15 massive stars, respectively, are found between $12$ and
$40~M_{\odot}$. A ~ $60~M_{\odot}$ star could be expected using a
standard initial mass function for both superbubbles and, if formed
with the rest of the cluster, it would already have exploded as an
SN. Thus, the kinematics, the size and the excess soft X-ray emission
of these SBs can be explained by a SN exploding in a wind-blown SB.

\acknowledgments

We thank F. De Colle for insightful discussions and acknowledge
financial support from the PAPIIT-UNAM (IA101413-2). PFV and ARG
acknowledge finantial support by CONACyT grants 167611 and 167625, and
DGAPA-PAPIIT grant IG100214.

{}

%%%%%%%%%%%%%%%%%%%%%%%%%%%%%%%%%%%%%%%%%%%%%%%%%%%%%%

\newpage

%%%%%%%%%%%%%%%%%%%%%%%%%%%%%%%%%%%%%%%%%%%%%%%%%%%%%%%%
\begin{deluxetable}{ccccccc}
\tabletypesize{\scriptsize}
\tablecaption{Log of Observations in this paper.
\label{obslog}}
\tablewidth{0pt}
\tablecolumns{2}
\tablehead{
\colhead{} &
\colhead{Target name} & \colhead{Obs ID} &
\colhead{Position (J2000)} & \colhead{Date} &\colhead{Time}& \colhead{Effective} \\
 & & & (RA, DEC) & (yyyy/mm/dd)&Exposure (ks)& Exposure (ks)
}
\startdata
& N\,70  & 0503680201 &
($05^{\rm h}43^{\rm m}25\fs0$, $-67^{\rm d}51^{\rm m}12\fs0$) &
2008/01/26&38.3 &21.0(MOS)/12.5 (pn)\\

& N\,185 & 0503680101 &
($04^{\rm h}53^{\rm m}47\fs40$, $-69^{\rm d}59^{\rm m}15\fs0$) &
2008/03/10&20.5&8.3 (MOS)/5.5 (pn)
\label{tab1}
\enddata
\end{deluxetable}
%%%%%%%%%%%%%%%%%%%%%%%%%%%%%%%%%%%%%%%%%%%%%%%%%%%%%%%%%%%

%%%%%%%%%%%%%%%%%%%%%%%%%%%%%%%%%%%%%%%%%%%%%%%%%%%%%%%%%%%%%%%%%%%%%%%%%%%%%%%
%%%%%%%%%%%%%%%%%%%%%%%%%%%%%%%%%%

\begin{table}[h]\footnotesize
\centering
 \caption{X-ray from N\,70:Best-fit parameters of the spectral model}
  \begin{tabular}{ccccccccccll}
   \hline
& & \multicolumn{3}{c}{}\\
\cline{5-6}
Model& $N_{H}$ &$k\,T$ &$n_{\rm e}n_{\rm p}V$\\
&$\times 10^{20}$ cm$^{-2}$& keV&$\times 10^{58} {\rm cm}^{-3}$\\

Absorption&5.0&-&\\
Apec
& &0.22(0.21-0.23)&3.4$\pm$0.1\\
\hline
\hline
\multicolumn{2}{l}{Flux=$(0.7 \pm 0.1)\times 10^{-12}$  erg cm$^{-2}$ s$^{-1}$}\\

\hline
\multicolumn{2}{l}{$\chi^{2}/d.o.f=320.57/242$}\\

\hline
\hline

\label{tab2}

\end{tabular}
\end{table}

%%%%%%%%%%%%%%%%%%%%%%%%%%%%%%%%%%%%%%%%%%%%%%%%%%%%%%%%%%%%%%%%%%%%%%%%%%%

%%%%%%%%%%%%%%%%%%%%%%%%%%%%%%%%%%%%%%%%%%%%%%%%%%%%%%%%%%%%%%%%%%%%%%%%%%%
\begin{table}[h]\footnotesize
\centering
 \caption{X-ray from N\,185:Best-fit parameters of the spectral model}
  \begin{tabular}{ccccccccccccll}

   \hline
& & \multicolumn{3}{c}{}\\
\cline{5-6}
Model& $N_{H}$ &$k\,T$&$n_{\rm e}n_{\rm p}V$\\
&$\times 10^{20}$ cm$^{-2}$& keV&$\times 10^{58} {\rm cm}^{-3}$\\

Absorption&5.0&-&\\
Apec& &0.20(0.19-0.21)&3.2$\pm$0.4\\
\hline
\hline
\multicolumn{2}{l}{Flux=$(0.6\pm 0.2) \times 10^{-12}$  erg cm$^{-2}$ s$^{-1}$}\\

\hline
\multicolumn{2}{l}{$\chi^{2}/d.o.f=151.02/94$}\\

\hline
\hline

\label{tab3}

\end{tabular}
\end{table}

%%%%%%%%%%%%%%%%%%%%%%%%%%%%%%%%%%

%%%%%%%%%%%%%%%%%%%%%%%%%%%%%%%%%%%%%%%%%%%%%%%%%%%%%%%%%%%
\begin{table}
\centering
 \caption{Soft X-ray luminosities, comparison between observed and predicted of the SBs}
  \begin{tabular}{lccc}
  \hline
& XMM-Newton observation& Predicted&Numerical\\
& & (Weaver at al.)& model\tablenotemark{*}\\
& $L_\mathrm{X}$ ($10^{35}~\mathrm{erg~s^{-1}}$)& $L_\mathrm{X}$ ($10^{35}~\mathrm{erg~s^{-1}}$)& $L_\mathrm{X}$ ($10^{35}~\mathrm{erg~s^{-1}}$)\\
N 70 & $2.4 \pm 0.4 $& $0.6$&$2.0$ \\
N 185 & $2.1 \pm 0.7$ & $0.05$ \\
 \hline
\label{tab4}
\end{tabular}
\end{table}

%%%%%%%%%%%%%%%%%%%%%%%%%%%%%%%%%%%%%%%%%%%%%%%%%%%%%%%%%%%%%%%%%%%%%%%%%%%%%%%

\end{document}